\documentclass[pra,twocolumn,notitlepage,superscriptaddress,showkeys,showpacs]{revtex4-1}
\usepackage{amsmath,amsfonts,amssymb,amsthm,epsfig,array}
\usepackage{dsfont}
\usepackage{graphics}
\usepackage{float}
\usepackage{verbatim}
\usepackage[dvipsnames]{xcolor}
\usepackage[hidelinks,colorlinks,linkcolor=blue,
citecolor=blue,urlcolor=blue]{hyperref}
\usepackage[titletoc,title]{appendix}

\newcommand{\bsl}[1]{\boldsymbol{#1}}

\newcommand{\ii}{\mathrm{i}}

\newcommand{\eqnref}[1]{Eq.\,\eqref{#1}}
\newcommand{\figref}[1]{Fig.\,\ref{#1}}

\newcommand{\mat}[1]{\left(\begin{matrix}#1\end{matrix}\right)}

\newcommand{\eq}[1]{\begin{equation} #1 \end{equation}}
\newcommand{\eqn}[1]{\begin{eqnarray} #1 \end{eqnarray}}
\let\oldAA\AA
\renewcommand{\AA}{\text{\normalfont\oldAA}}
\newcommand{\sgn}[1]{\text{sgn}(#1)}

\newcommand{\si}{\sigma}

\newcommand{\mbt}{{MnBi$_2$Te$_4$}}

\newcommand{\eqJA}{Eq.\,({\color{blue}6})}
\newcommand{\eqS}{Eq.\,({\color{blue}4})}
\newcommand{\eqAFMR}{Eq.\,({\color{blue}13})}

\begin{document}
\title{Magnetic-Resonance-Induced Pseudo-electric Field and Giant Current Response in Axion Insulators}
\author{Jiabin Yu}
\affiliation{Department of Physics, the Pennsylvania State University, University Park, PA, 16802}
\author{Jiadong Zang}
\affiliation{Department of Physics and Astronomy, University of New Hampshire, Durham, NH 03824}
\author{Chao-Xing Liu}
\email{cxl56@psu.edu}
\affiliation{Department of Physics, the Pennsylvania State University, University Park, PA, 16802}
\begin{abstract}
A quantized version of the magnetoelectric effect, known as the topological magnetoelectric effect, can exist in a time-reversal invariant topological insulator with all its surface states gapped out by magnetism.
This topological phase, called the axion insulator phase,
has been theoretically proposed but is still lack of conclusive experimental evidence due to the small signal of topological magnetoelectric effect.
In this work, we propose that the dynamical in-plane magnetization in an axion insulator can generate a ``pseudo-electric field", which acts on the surface state of topological insulator films and leads to the non-zero response current.
Strikingly, we find that the current at magnetic resonance (either ferromagnetic or anti-ferromagnetic) is larger than that of topological magnetoelectric effect by several orders of magnitude, and thereby serves as a feasible smoking gun to confirm the axion insulator phase in the candidate materials.
\end{abstract}
\maketitle


{\it Introduction: }
More than forty years ago, the axion was proposed as an elementary particle to resolve the strong CP problem in high-energy physics~\cite{Peccei1977AxionPRL,Peccei1977AxionPRD,
Weinberg1978Axion,Wilczek1978Axion}.
Later studies suggest that the axion might be a candidate for the dark matter in the universe~\cite{Preskill1983Axion,Abbott1983Axion,Dine1983AxionDM,Ipser1983Axion}.
While axions so far remain experimentally elusive, it has been proposed that the electrodynamics of axions~\cite{Wilczek1987Axion} may effectively exist in a variety of solid state systems, in particular the system based on the topological insulator (TI)~\cite{Hasan2010TI,Qi2010TITSC,Qi2008TFT}.
In contrast to the conventional Maxwell's equations for a trivial insulator, the electromagnetic response in the bulk of TIs requires an additional term (known as the $\theta$ term) in the action:
\eq{
\label{eq:S_theta}
S_\theta=\frac{e^2}{hc} \int dt d^3 r \frac{\theta}{2\pi} \bsl{E}\cdot \bsl{B}\ ,
}
where $e$ is the elementary charge and $\theta$ is the dimensionless pseudoscaler axion field.
If time reversal (TR) symmetry is preserved, $\theta$ can only take two topologically distinct values in the bulk of a system: $0$ for a trivial insulator and $\pi$ for a TI.
The gauge transformation can change the value of $\theta$ by $2\pi n$ with $n$ an arbitrary integer without affecting the bulk topology, reflecting the $Z_2$ topological classification.
As a consequence, gapless modes must exist at the interface between a TI and a trivial insulator (or the vacuum) in the presence of TR symmetry, as $\theta$ cannot vary continuously without gap closing or TR-breaking effects.~\cite{Qi2008TFT}
This topological surface state leads to a variety of exotic phenomena in TI materials, including the quantum anomalous Hall (QAH) effect~\cite{Chang2013QAH}, the topological magneto-optical effect~\cite{Wu2016THzTI,Dziom2017THzTI,Okada2016THzQAH} even with exact quantization in terms of fine-structure constant~\cite{Wu2016THzTI}, topological magnetoelectric effect (TME)~\cite{Qi2008TFT,Wang2015AI} and the image magnetic monopole~\cite{Qi2009AIMono}.

When {\it all} the surface modes of a TR-invariant TI are gapped out by the surface magnetic coating with a hedgehog magnetization configuration, the $2\pi n$ ambiguity can be removed and the $\theta$ value is uniquely determined~\cite{Qi2008TFT,Qi2009AIMono,Essin2009AI,Wang2015AI,Morimoto2015AI}.
This system with a well-defined non-zero $\theta$ field in \eqnref{eq:S_theta} is defined as the axion insulator (AI)~\cite{Wang2015AI,Mogi2017AI,Xiao2018AI}.
The polarization (magnetization) of an AI can be induced by a magnetic (electric) field in the same direction with the response coefficient quantized to $\frac{\theta e^2}{2\pi h c}$~\cite{Qi2008TFT,Wang2015AI}, serving as a conclusive experimental signature to distinguish an AI from a trivial insulator.
Such effect is called TME and requires $\theta$ to be well-defined everywhere in the system since $\theta$ determines the experimentally measurable magnetoelectric coefficient.
Besides the unique TME, the AI also exhibits the zero Hall resistance with large longitudinal resistance, which is nevertheless not conclusive since it can also happen in trivial insulators.
The AI phase has been proposed in the ferromagnetic insulator-TI-ferromagnetic insulator (FMI-TI-FMI)  heterostructure~\cite{Wang2015AI,Mogi2017AI,Xiao2018AI}, anti-ferromagnetic topological insulator \mbt~\cite{Otrokov2018MnBi2Te4,Gong2018MnBi2Te4,Li2018MnBi2Te4,Zhang2018AIMnBi2Te4} and various other materials~\cite{Li2010DAF,Mogi2017AIMTI,Chowdhury2019MBSeAI,Gui2019EIPAI,Xu2019EIAAI}.
Although the zero Hall plateau has been observed in the FMI-TI-FMI heterostructure~\cite{Mogi2017AI,Xiao2018AI}, the conclusive TME has not been detected due to the small magnetoelectric current.
Therefore, identifying a testable transport signal to distinguish the AI from a trivial insulator is the major challenge of the field.

\begin{figure}[t]
\includegraphics[width=\columnwidth]{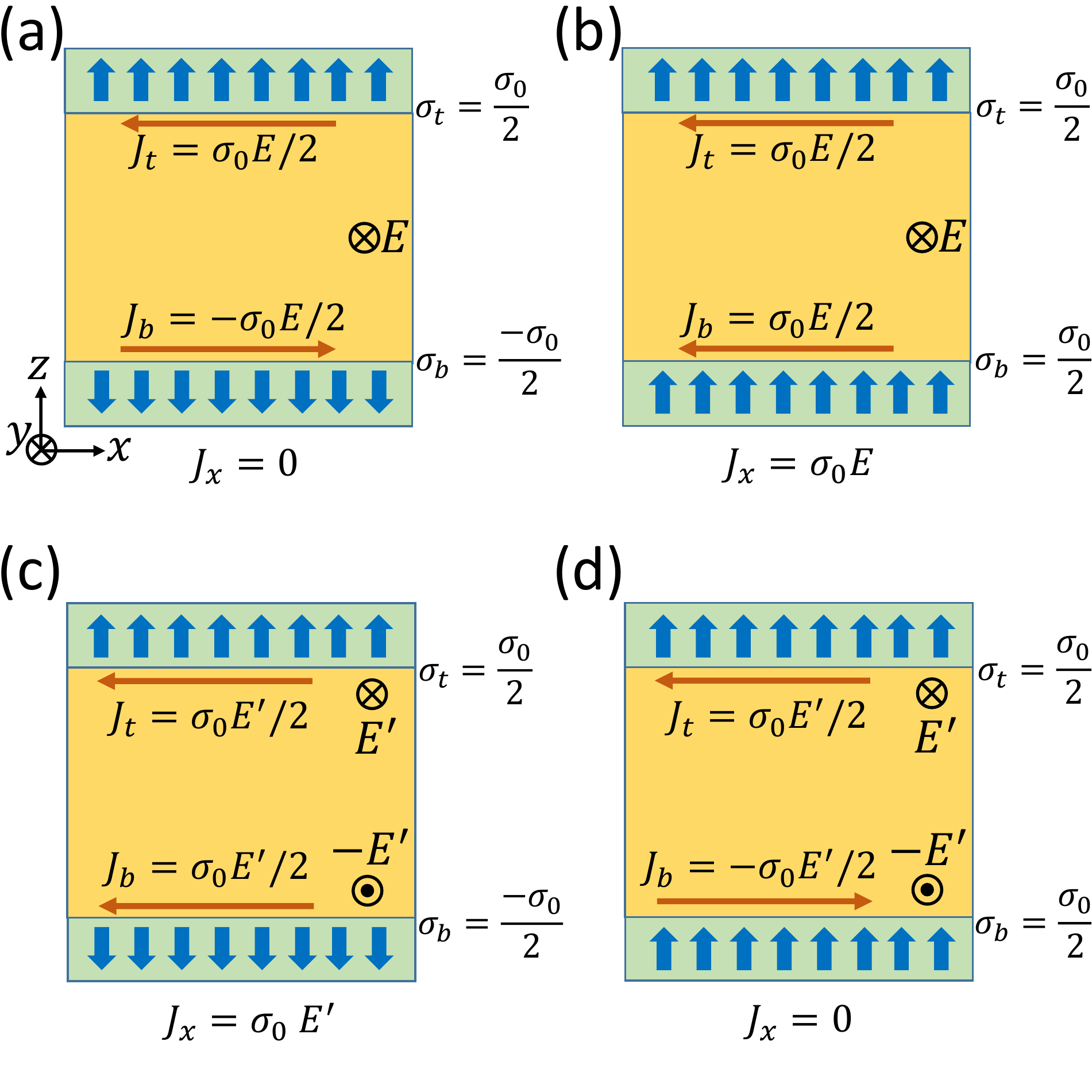}
\caption{\label{fig:setup_AI}
In the four graphs, the yellow middle parts are TIs, and the green parts are FMIs with blue arrows the magnetization.
(a) and (c) are AIs with opposite surface Hall conductance $\sigma_t=-\sigma_b$, and (b) and (d) are QAH insulators with $\sigma_t=\sigma_b$.
In the uniform electric field $E$ along $y$, the AI in (a) has zero total Hall current ($J_x=0$) due to the opposite surface Hall currents ($J_t=-J_b$), while the QAH insulator in (b) has non-zero Hall current with Hall conductance $\sigma_0=-e^2/h$.
If the electric field $E'$ is in opposite $y$ directions on the two surfaces, AI in (c) has non-zero Hall current, but the QAH insulator in (d) has zero Hall current.
}
\end{figure}

In this letter, we propose that the magnetic resonance (MR) in an AI can induce a pseudo-electric field (PEF), leading to a response current which is much larger than that of TME.
This current response cannot exist in a trivial insulator and thus serves as a feasible and unequivocal experimental evidence to identify the AI phase.
Moreover, our proposal serves as the first concrete example of MR-induced PEF in realistic materials.
The intuitive picture is summarized in \figref{fig:setup_AI}, which takes FMI-TI-FMI heterostructure as an example.
In the FMI-TI-FMI heterostructure, the QAH state can exist when the two FMIs have parallel magnetic moments (\figref{fig:setup_AI}(b),(d)), while the AI phase is expected for the anti-parallel configuration (\figref{fig:setup_AI}(a),(c)).
The TI surface states open a gap due to magnetic proximity and show half quantized Hall conductance with its sign depending on the magnetization direction.
With a uniform electric field, the Hall currents of the top and bottom surfaces have the same direction in the QAH phase (\figref{fig:setup_AI}(b)) but cancel each other in \figref{fig:setup_AI}(a), leading to the zero Hall plateau of the AI phase.
However, zero Hall plateau may also occur in a trivial or Anderson insulator~\cite{Feng2015ZHP,Wang2014QAHPT,Chang2016QAHAI,Chen2019ZHP}.
In contrast, for an in-plane electric field with opposite directions at two surfaces, the Hall current is expected to be non-zero in the AI phase (\figref{fig:setup_AI}(c)) but vanishes for the QAH state (\figref{fig:setup_AI}(d)).
As a direct consequence of TME in AIs~\cite{Wang2015AI}, the non-uniform electric field and the resultant current response in \figref{fig:setup_AI}(c) can be generated by a time-dependent magnetic field, but the current magnitude is limited by the TI film thickness (maximally tens of nanometers).
Instead of electric fields, we consider the dynamics of in-plane magnetization in the FMI layers.
The in-plane magnetization acts on the TI surface states effectively as a time-dependent pseudo-gauge field (PGF) \cite{Iorio2015PGF}, and thus generates a PEF of the same form as the physical electric field in \figref{fig:setup_AI}(c) and (d), leading to non-zero current response in the AI phase.
In particular, our estimation shows that the current induced by the PEF at ferromagnetic resonance (FMR) in the heterostructures (or the antiferromagnetic resonance (AFMR) in the {\mbt} system) is giant.

{\it PEF induced by dynamical in-plane magnetization: }
We start from demonstrating that the dynamical in-plane magnetization of the FMIs in \figref{fig:setup_AI} can induce the PEF and the current response.
The low-energy physics of the FMI-TI-FMI heterostructure is given by the surface states of the TI film coupled to surface magnetization and the external electromagnetic field, resulting in the following Hamiltonian~\cite{Garate2010ISGE}:
\eq{
H=\int \frac{d^2 k}{(2\pi)^2}\sum_{i} c^\dagger_{\bsl{k},i} [h_{0,i}+h_{Z}+h_{ex,i}] c_{\bsl{k},i}\ .
}
\eq{
h_{0,i}=v_{f,i}[- \sigma_y (\hbar k_x+\frac{e}{c}A_{i,x})+ \sigma_x (\hbar k_y+\frac{e}{c}A_{i,y})]+(-e) \varphi_i
}
depicts the surface Dirac modes $c^\dagger_{\bsl{k},i}=(c^\dagger_{\bsl{k},i,\uparrow},c^\dagger_{\bsl{k},i,\downarrow})$ coupled to the the 2+1D physical gauge field $A^\mu_i=(\varphi_i,A_{i,x},A_{i,y})$, where $i=t,b$ labels the top and bottom surfaces, respectively, $\bsl{k}=(k_x,k_y)$, $v_{f,t}=-v_{f,b}=v_f$ and $\sigma_{x,y,z}$ are Pauli matrices for spin.
$h_{Z}=\mu_B \bsl{B}\cdot \bsl{\si}$ is the Zeeman term with $\mu_B$ Bohr magneton and $\bsl{B}$ the uniform magnetic field, and $h_{ex}=g_M \bsl{M}_i\cdot \bsl{\si}$ is the exchange coupling term with $g_M$ assumed to be positive and the same on both surfaces for simplicity.
We notice that the in-plane components of the Zeeman and exchange coupling terms play the same role as the vector potential and thus can be regarded as the PGF.
For convenience, we transform the creation operator to the Grassmann field $\bar{\psi}_{i}$ and rewrite the Hamiltonian into the action form
\eq{
\label{eq:S}
S=\int d^3 x \bar{\psi}_{i}[\Gamma^\mu_i(\ii \hbar  \partial_\mu-\frac{e}{c} \widetilde{A}_{i,\mu})- \frac{m_{i,z}}{c}\sigma_z ]\psi_{i}
}
where $x^\mu=(c t,x,y)$ and $\Gamma^\mu_i=(1,-\frac{v_{f,i}}{c}\sigma_y,\frac{v_{f,i}}{c}\sigma_x)$.
In \eqnref{eq:S}, $m_{i,z}$ plays the role of mass, and $\widetilde{A}_{i,\mu}=A_{i,\mu}+A^{pse}_{i,\mu}$ contains the PGF
\eq{
\label{eq:A_pse}
A^{pse}_{i,\mu}=\frac{c}{e v_{f,i}}(0, - m_{i,y}, m_{i,x} )\ ,}
where $\bsl{m}_{i}=\mu_B \bsl{B}+g_M \bsl{M}_{i}$.
The corresponding ``electric" field of $\widetilde{A}_{i,\mu}$ can be written as $\widetilde{E}_{i,a}=-\partial_{a}\varphi_i-\frac{1}{c}\frac{\partial \widetilde{A}_{i,a}}{\partial t}=E_{i,a}
+E^{pse}_{i,a}$ with $E_{i,a}$ the conventional electric field and $a=x,y$.
We call $E^{pse}_{i,a}=\frac{1}{e v_{f,i}}( \dot{m}_{i,y}, - \dot{m}_{i,x} )$ the PEF, following the terminology used for pseudo-magnetic field induced by strain in graphene \cite{Levy2010StrainPMF}.
Next we derive the response current generated by PEF based on \eqnref{eq:S}.

By integrating out the fermionic modes in \eqnref{eq:S}, the response of the system to the leading order can be obtained:
\begin{equation}
\label{eq:J_A}
J^{\mu}_{i}=\sigma_{i}\varepsilon^{\mu\rho\nu}\partial_\rho\tilde{A}_{i,\nu}\ ,
\end{equation}
where $J^{\mu}_{i}=(c \rho_i, J^x_i, J^y_i)$ is the current density of $i$ surface and
$\partial_\rho=\partial_{x^\rho}$.
\begin{equation}
\label{eq:HC}
\sigma_{i}=-\text{sgn}(m_{i,z})\frac{e^2}{2h}
\end{equation} is the Hall conductance of $i$ surface, showing that the surface Hall conductance is determined by the sign of the surface gap.
We now focus on the AI phase with anti-parallel magnetization alignment.
As the exchange coupling is generally much larger than Zeeman coupling ($|g_M M_{i,z}|\gg |\mu_B B_z|$), we expect opposite Hall conductance on two surfaces ($\sigma_{b}=-\sigma_t$). Therefore, the total current density only depends on the difference between $\widetilde{A}_{t}$ and $\widetilde{A}_{b}$ as
\eq{\label{eq:J_AP}
J^\mu_{AP}=\sigma_t\varepsilon^{\mu\rho\nu}\partial_\rho (\tilde{A}_{t,\nu}-\tilde{A}_{b,\nu})\ .
}
Thus, the PEF can induce currents in the same way as the physical electric field according to $\varepsilon^{\mu\rho\nu}\partial_\rho \widetilde{A}_{i,\nu}=(\widetilde{B}_{i,z}, \widetilde{E}_{i,y}, -\widetilde{E}_{i,x})$, and the physics in \figref{fig:setup_AI}(a) and (c) can be described by \eqnref{eq:J_AP} if choosing
$E=(\widetilde{E}_{t,y}+\widetilde{E}_{b,y})/2$ and $E'=(\widetilde{E}_{t,y}-\widetilde{E}_{b,y})/2$.
In the following, we consider a simple case where the uniform magnetic field only has an oscillating $x$ component, {\it i.e.} $\bsl{B}(t)=(B_0 \cos(\omega t),0,0)$ with the constant $B_0$, in order to estimate the current magnitude.

The oscillating uniform magnetic field can induce a non-uniform electric field along $y$ owing to the Faraday's law: $E_y(t,z)=-\omega B_0 \sin(\omega t) z/c$ with $z=0$ set at the middle of the TI layer.
In this case, the physical gauge field in \eqnref{eq:S} must satisfy
$\varepsilon^{1\rho\nu}\partial_\rho A_{i,\nu}=E_y(t,z_i)=-\omega B_0 \sin(\omega t) z_i/c$,
where $z_{t (b)}=(-) L_z/2$ and $L_z$ is the distance between two surfaces.
In addition, $\bsl{B}$ can also drive the surface magnetic moments away from the $z$ direction and thus induce the time-dependent in-plane magnetization $M_{i,a}$.
In sum, under the adiabatic approximation $\hbar \omega\ll
|g_s M_{i,z}|$, we have
\eq{
\label{eq:J_x_AP}
J^x_{AP} =J_E + J_Z + J_M\ ,
}
for the anti-parallel case.
In the above equation, $J_E=-\frac{1}{c}\sigma_t B_0 \omega \sin(\omega t) L_z$ is the TME current density, $J_Z=2 \si_t  E^{pse}_Z$ with $E^{pse}_Z=\frac{1}{e v_f} \mu_B B_0 \omega \sin(\omega t)$ the PEF induced by the in-plane Zeeman term, and $J_M=2 \si_t  E^{pse}_M$ with $E^{pse}_M=-\frac{1}{2 e v_f} g_M(\dot{M}_{t,x}+\dot{M}_{b,x})$.
Among these contributions, let us first estimate $J_E$ and $J_Z$.
With typical values of parameters $L_z = 20$nm, $v_f= 6.5 \times  10^5$m/s, $B_0=10$G and $L_y=200 \mu$m (the length of the sample along $y$)~\cite{Zhang2018AIMnBi2Te4}, the current amplitude of TME is estimated as (after converting to SI unit) $I_{E}=|\text{max}(J_E) L_y|=0.5 (\frac{\omega}{2\pi \text{GHz} }) \text{nA}$, which is small for GHz frequency.
On the other hand, the current induced by the Zeeman effect can be neglected as $|J_Z/J_{E}|\approx 9\times 10^{-3}$.
In the next section, we focus on the current response generated by the magnetization-induced PEF, {\it i.e.} $J_M$.

\begin{figure}[t]
\includegraphics[width=\columnwidth]{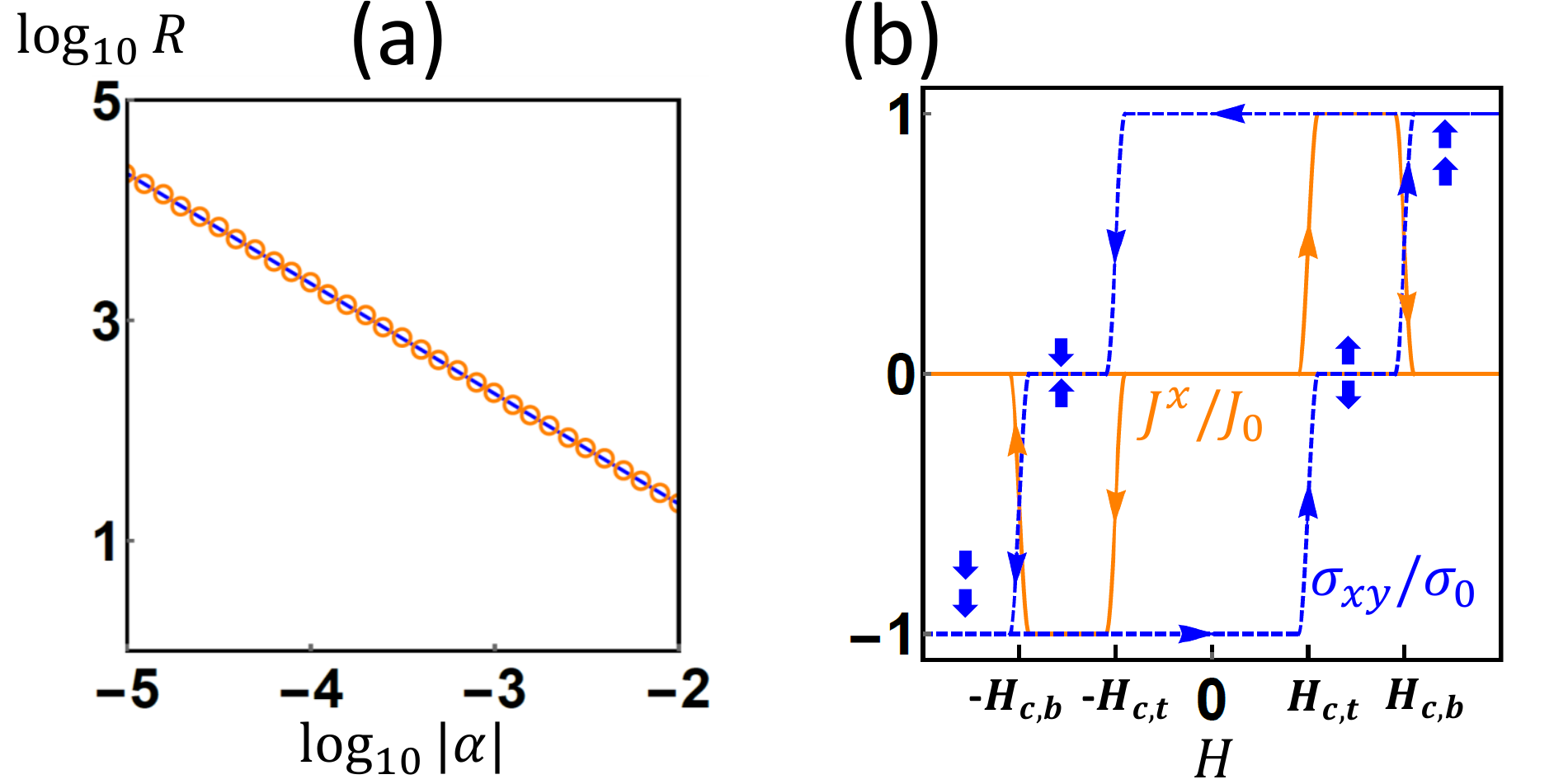}
\caption{\label{fig:Jx}
(a) shows how the ratio between the FMR-induced and TME currents,  noted as $R$, changes with the damping constant $\alpha$.
The orange circle (blue line) is given by the solution of LLG equation with (without) higher-order terms of $\alpha$.
(b) schematically shows how the FMR-induced {\it AC} current density along $x$  ($J^x$ in orange) and the total {\it DC} Hall conductance $\sigma_{xy}$ (blue) change with the initializing magnetic field $H$ along $z$. Here the setup considered is shown in \figref{fig:setup_AI} and the current is measured after decreasing $H$ to guarantee FMR on both surfaces.
The on-line arrows indicate the direction of changing $H$, while the vertical off-line blue arrows imply the magnetization configuration at the corresponding plateau of the blue line.
$J_0$ is the postfactor of $\text{sgn}(M_{t,z})$ in \eqnref{eq:J_M_FM}.
}
\end{figure}

{\it FMR in FMI-TI-FMI heterostructures:}
In the FMI-TI-FMI heterostructure, the in-plane magnetization and the induced current are maximized at the FMR frequency of the FMI layer.
Since two FMI layers are decoupled by the TI layer, the Landau-Lifshitz-Gilbert (LLG) equations~\cite{Kittel1948FMR,Kittel1976SSP,Aharoni2000FM} for two FMI layers under the same uniform magnetic field $\bsl{B}(t)$ have the same form and can be treated separately.
The equation is solved in the limit that the second and higher order terms of $|M_{i,x}|/|\bsl{M}_i|$, $|M_{i,y}|/|\bsl{M}_i|$ and $|B_0|/|\bsl{M}_i|$ are small enough to be neglected, and the steady solution of $M_{i,x}$ at FMR is given by
\begin{equation}
\label{eq:M_FMR}
M_{i,x}
=
\frac{\gamma_0 B_0 M_s }{2 \alpha \omega_0}
\sin(\omega_0 t)\ ,
\end{equation}
where $\alpha$ is the dimensionless damping constant, $M_s=|\bsl{M}_i|$ is the magnetization amplitude, $\omega_0$ is resonance frequency and $\gamma_0=2 e/(2 m_e c)$ is the magneto-mechanical ratio of electrons (see Sec.\,B in~\cite{SMs}).
In the derivation of the above expression, we assume the same magnetization amplitude, resonant frequency and damping constant for two FMIs, and only keep the leading order term of $\alpha$.
Since two FMIs are driven by the same uniform magnetic field, $M_{t,x}$ and $M_{b,x}$ have the same sign in \eqnref{eq:M_FMR}.
The typical FMR frequency is $\omega_0=2\pi$GHz, and its energy scale $\hbar \omega_0\approx 4\mu$eV is much smaller than the magnetic gap of the FMI ($|g_M M_s|= 0.1 meV$)~\cite{Xiao2018AI}.
Thus the adiabatic approximation holds, and we can combine \eqnref{eq:M_FMR} with \eqnref{eq:J_x_AP} to get the current density induced by magnetic dynamics at the FMR:
\eq{
\label{eq:J_M_FM}
J_M=\sgn{M_{t,z}} \frac{e B_0 \gamma_0 g_M M_s}{2 v_f h \alpha} \cos(\omega_0 t)\ .
}
Then the ratio between the amplitudes of $J_M$ and $J_E$ is
\eq{
\label{eq:R_FM}
R=\left|\frac{\max(J_M)}{\max(J_E)}\right|=\left|\frac{\hbar/L_z}{m_e v_f}\right| \left|\frac{g_M M_s}{\hbar \omega_0}\right|\frac{1}{|\alpha|}\approx \frac{0.2}{|\alpha|}
}
with the TME current $I_E=0.5$nA.
Although the above equation is obtained by neglecting higher-order terms of $\alpha$, the approximation is quite good as shown in \figref{fig:Jx}(a).
In a typical range $|\alpha|=10^{-5}\sim 10^{-2}$ for FMIs \cite{Hauser2016YIGFMR}, the ratio is approximately $R\approx20\sim 2\times 10^4$, resulting in the experimentally measurable current amplitude $I_M= I_E R\approx 10\sim 10^4$nA.
Therefore, the current response induced by magnetic dynamics at FMR is the dominant contribution, {\it i.e.} $J^x_{AP}\approx J_M$ for \eqnref{eq:J_x_AP}, and can be used to distinguish the AI from a trivial insulator experimentally.

We next compare the current response induced by FMR to the standard {\it dc} Hall conductance in the FMI-TI-FMI heterostructures when varying initializing magnetic fields.
Experimentally, the FMI-TI-FMI heterostructure is realized by inserting a TI layer between a Cr-doped TI layer (top) and a V-doped TI layer (bottom).~\cite{Xiao2018AI}
Since the coercive field $H_{c,t}$ of Cr-doped layer is around $0.14$T, much smaller than $H_{c,b}\sim 1$T of V-doped layer, a two-step transition of Hall conductance, schematically shown by the dashed blue line in \figref{fig:Jx}(b), has been demonstrated in experiments (see Fig.\,2 in Ref.\cite{Xiao2018AI}).
The AI phase is expected to exist when the Hall conductance is zero with anti-parallel magnetization at two surfaces in the intermediate field ranges $-H_{c,b}<H<-H_{c,t}$ and $H_{c,t}<H<H_{c,b}$.
When the state with zero Hall conductance is achieved, the mechanism discussed here will induce a large current response at the FMR frequency.
We emphasize that the initializing magnetic field should be reduced or removed before measuring the current response of FMR to guarantee a similar FMR frequency of two FMI layers.~\footnote{
It is because the initializing field $H$ can change the resonant frequencies of two FMIs.~\cite{Kittel1948FMR} If two FMIs have the same (slightly different) zero-field resonant frequencies, $H$ should be reduced to zero (a small nonzero value to compensate the difference) before measuring the FMR-induced current.}
On the other hand, when the {\it dc} transport measurement shows a QAH state with Hall conductance $\sigma_{xy}=\pm e^2/h$, the current response at the FMR frequency is expected to be quite small owing to the opposite directions of FMR-induced PEFs on the two surfaces.~\cite{SMs}
The behaviors of {\it dc} transport and the current measurement at the FMR frequency are schematically shown in \figref{fig:Jx}(b), and the sharp contrast between these two measurements can serve as the key evidence of AI phase.

{\it  AFMR in {\mbt}: }
{\mbt} has A-type anti-ferromagnetism (AFM): ferromagnetic layers with opposite out-of-plane magnetization are alternatively stacked along the $z$ direction.
Due to the combined symmetry of half translation and TR for AFM, the bulk Hamiltonian of {\mbt} is the same as the TI Hamiltonian of Bi$_2$Te$_3$~\cite{Zhang2018AIMnBi2Te4}.
The topological surface states on both surfaces are gapped by ferromagnetic layers, resulting that the low-energy action of {\mbt} has the same form as \eqnref{eq:S}.
Due to the intrinsic magnetism in {\mbt}, the exchange coupling between surface electrons and magnetization is much stronger than that of the proximity effect in the FMI-TI-FMI heterostructure and leads to a larger magnetic gap ($g_M M_s\approx 0.1$eV) of surface states~\cite{Otrokov2018MnBi2Te4,Gong2018MnBi2Te4,Li2018MnBi2Te4,Zhang2018AIMnBi2Te4}.
In the following, we consider an even number of layers of {\mbt} films so that the top and bottom layers have anti-parallel magnetization.
To describe the magnetic dynamics of AFM in {\mbt}, particularly around the AFMR, the exchange interaction of magnetization between the neighboring layers should be included in the LLG equation and leads to a larger resonance frequency $\omega_1\sim$THz~\cite{Kittel1951AFMR, Keffer1952AFMR, Wang2018AFMR}.
Since $0.1 eV\gg h$(1THz)$\approx 4 $meV, the adiabatic approximation is still valid and \eqnref{eq:J_x_AP} can be applied in this case.

The LLG equation for this AFM system can be solved with the same approximation as the FMR case, and the steady solution at AFMR reads
\eq{
\label{eq:M_AFMR}
M_{i,x}
=
\frac{\gamma_0 B_0 M_s }{2 \alpha \omega_1}\frac{B_A}{B_A+B_E}
\sin(\omega_1 t)
\ ,}
where $B_E$ and $B_A$ are the exchange field and anisotropy field, respectively. (See details in Sec.\,C of \cite{SMs}.)
The resulting current $J_M$ from \eqnref{eq:M_AFMR} is derived as
\eq{
J_M=\sgn{M_{t,z}} \frac{e B_0 \gamma_0 g_M M_s}{2 v_f h \alpha} \frac{B_A}{B_A+B_E}\cos(\omega_1 t)\ .
}
By choosing $g_M M_s=0.1 $eV, $\omega_1=2 \pi $ THz and all other parameters the same as the FMR case,
we find that the current $J_Z$ induced by Zeeman coupling is still negligible, while the TME current amplitude becomes $I_E=500$nA owing to the increase of the resonance frequency.
The ratio between the amplitudes of $J_M$ and $J_E$ now reads
\eq{
\label{eq:R_FM}
R=\left|\frac{\max(J_M)}{\max(J_E)}\right|=\left|\frac{\hbar/L_z}{m_e v_f}\frac{g_M M_s}{\hbar \omega_1}\frac{B_A}{B_A+B_E}\right|\frac{1}{|\alpha|}\approx \frac{0.2}{|\widetilde{\alpha}|}
\ ,}
where $\widetilde{\alpha}=\alpha (B_A+B_E)/B_A$.
By choosing a typical ratio between the exchange and anisotropy fields $|B_E/B_A|=10^2$~\cite{Keffer1952AFMR} and the same typical range of $|\alpha|$ as the FMR case, we find $R\approx 0.2\sim 200$. Thus, the AFMR-induced current may still be dominated when $|\alpha|$ can be reduced, and its amplitude ($I_M\approx 0.1 \sim 100\mu A$) is much larger than the FMR case.
Since the magnetization along $x$ has the same form on two surfaces according to \eqnref{eq:M_AFMR}, the AFMR-induced current in the QAHI phase (odd number of layers) of {\mbt} is zero, similar as the FMR case. This suggests an even-odd effect of the AFMR-induced current response in {\mbt} films due to different surface magnetization configurations.

{\it Conclusion and Discussion: }
In summary, we have demonstrated that magnetic dynamics in the FMI-TI-FMI heterostructure and {\mbt} can give rise to PEF, which in turn generates a giant current response at the FMR or AFMR in AIs but not in trivial insulators or QAH insulators.
Given the observation of zero Hall plateau~\cite{Xiao2018AI,Mogi2017AI}, this phenomenon awaits for the experimental test in FMI-TI-FMI heterostructure.
Current experiments on {\mbt} films have shown heavy electron-doping~\cite{Otrokov2018MnBi2Te4}, which is detrimental to the mechanism proposed here.
Therefore, an electric gate is required on {\mbt} films and our theory predicts that the AFMR-induced current response will be greatly enhanced when the Fermi energy is gated into the magnetic gap.
Although the PEF has been studied in graphene with the dynamical strain~\cite{Vaezi2013StrainGraphene}, our proposal of MR-induced PEF is more feasible since MR has been observed and studied since 1940s~\cite{Griffiths1946FMR,Yager1947FMR,Kittel1948FMR}.
Our AFMR-induced current has a fundamentally different mechanism from that induced by the bulk dynamical axion field discussed in Ref.~\cite{Li2010DAF,Sekine2016DAF}, as the latter requires a non-zero external static magnetic field that is absent in our proposal.
Our theory unveils the intriguing interplay between magnetic dynamics and magnetoelectric response in the AI phase and will pave the way to a new class of electric-field-tunable axion devices for spintronics applications.~\cite{Sekine2016ElecAFMR}

{\it Acknowledgement: }
We acknowledge helpful discussions with Moses H. W. Chan, Cui-Zu Chang, Chao-Ming Jian, Nitin Samarth, Akihiko Sekine, Di Xiao and Peter Armitage.
The theoretical framework and concepts developed in this work are mainly supported by DOE grant (DE-SC0019064).
We also acknowledge the support of the Office of Naval Research (Grant No. N00014-18-1-2793) and Kaufman New Initiative research grant of the Pittsburgh Foundation.

\appendix
\begin{widetext}
\section{Details on Linear Response}
In this section, we derive the linear response formula {\eqJA} from the action in the main text.
The metric used here is $(-,+,+)$.

In order to derive the response, we integrate out the fermion modes and obtain the effective action of $\widetilde{A}_i$.
Note that {\eqS} in the main text does not contain coupling between the top and bottom surfaces, and thereby it can be written as $S=S_t+S_b$ with $S_i=S_{i,\psi}+S_{i,\psi A}$,
\begin{equation}
S_{i,\psi}=\int d^3 x \bar{\psi}_{i}(\ii \hbar \Gamma^\mu_i \partial_\mu - \frac{m_{i,z}}{c}\sigma_z )\psi_{i}\ ,\
S_{i,\psi A}=\int d^3 x \bar{\psi}_{i}(-\frac{e}{c}\Gamma^\mu_i \widetilde{A}_{i,\mu})\psi_{i}\ ,
\end{equation}
and $\Gamma^\mu_i=(1,-(v_{f,i}/c)\sigma_y,(v_{f,i}/c)\sigma_x)$.
The partition function can also be split into two parts $Z=Z_t Z_b$ with
\begin{equation}
Z_i=\int D\bar{\psi}_i D\psi_i e^{\frac{\ii}{\hbar} S_i}.
\end{equation}
The effective action of $\widetilde{A}_i$ can be derived from $Z_i$ by
\begin{equation}
 e^{\frac{\ii}{\hbar} S_{eff,i}[\widetilde{A}_i]}=Z_i/Z_i[\widetilde{A}_i=0]=\int D\bar{\psi}_i D\psi_i e^{\frac{\ii}{\hbar}(S_{i,\psi}+S_{i,\psi A})}\left./\int D\bar{\psi}_i D\psi_i e^{\frac{\ii}{\hbar}S_{i,\psi}}\right. \equiv \left\langle e^{\frac{\ii}{\hbar}S_{i,\psi A}}\right\rangle_{i,\psi}\ .
\end{equation}
As a result of the above equation, we have
\begin{equation}
\frac{\ii}{\hbar} S_{eff,i}[\widetilde{A}_i]=\sum_{n=1}^{\infty}\frac{1}{n!}\left\langle (\frac{\ii}{\hbar}S_{i,\psi A})^n\right\rangle_{i,\psi}^C\ ,
\end{equation}
where the upper index $C$ means to only include the connected graph.
The first order term ($n=1$) vanishes as
\begin{equation}
\left\langle \frac{\ii}{\hbar} S_{i,\psi A} \right\rangle _{i,\psi}^C=\frac{e}{\hbar c} \int d^3x \widetilde{A}_{i,\mu} (x) \int \frac{d^3 k}{(2\pi)^3} \text{Tr}[\Gamma^\mu_i G_{i}(k)]=0\ ,
\end{equation}
where $G_{i}(k)=[\Gamma^\mu_i k_\mu+\frac{m_{i,z}}{\hbar c}\sigma_z]^{-1}$ is the Green function and $k^{\mu}=(\omega/c,k_x,k_y)$.
The second order term ($n=2$) reads
\begin{equation}
\frac{1}{2!}\left\langle (\frac{\ii}{\hbar} S_{i,\psi A} )^2\right\rangle _{i,\psi}^C=\frac{\ii}{2 \hbar c^2}\int \frac{d^3 q}{(2\pi)^3} \widetilde{A}_{i,\mu}(-q)\widetilde{A}_{i,\nu}(q) f^{\mu \nu}_i(q)\ ,
\end{equation}
where
\begin{equation}
 f^{\mu \nu}_i(q)=\ii\frac{e^2}{\hbar}\int \frac{d^3 k}{(2\pi)^3} \text{Tr}[G_{i}(k)\Gamma^\mu_i G_{i}(k+q)\Gamma^\nu_i]\ ,
\end{equation}
and $\widetilde{A}_{i}^{\nu}(q)=\int d^3 x \widetilde{A}_{i}^{\nu}(x) e^{-i q x }$.
From
\begin{equation}
\label{eq:J_gen}
J^{\mu}_i=c^2 \frac{\delta S_{eff,i}[\widetilde{A}^{\mu}_i]}{\delta A_{i,\mu}}\ ,
\end{equation}
the current given by the $n\geq 3$ terms is of order $(A)^{n_1}(A^{pse})^{n_2}$ with $n_1+n_2\geq 2$.
Here we use the definition $\delta A_{i,\mu}(x)/\delta A_{i,\mu'}(x')=\delta_{\mu\mu'}\delta(x-x')$.
In spirit of linear response theory, which treats both $A$ and $A^{pse}$ as perturbation, we only include the lowest order contribution (linear response). 
Therefore, we neglect all $n\geq 3$ terms and the effective action reads
\begin{equation}
S_{eff,i}=\frac{1}{2 c^2}\int \frac{d^3 q}{(2\pi)^3} \widetilde{A}_{i,\mu}(-q)\widetilde{A}_{i,\nu}(q) f^{\mu \nu}_i(q).
\end{equation}
Since $S_{eff,i}$ should be gauge invariant under the gauge transformation $A_{i,\mu}\rightarrow A_{i,\mu}+\partial_\mu \Lambda_i$, all the gauge dependent terms are required to be zero.
We further neglect the contribution of second and higher derivatives of $A_i$ as they are small, and the effective action eventually reads
\begin{equation}
S_{eff,i}=\frac{\sigma_i}{2 c^2}\int d^3 x \epsilon^{\mu\rho\nu} \widetilde{A}_{i,\mu}\partial_\rho \widetilde{A}_{i,\nu}\ ,
\end{equation}
where
$
\sigma_i=-\sgn{m_{i,z}}e^2/(2h)
$
indicates the half quantized Hall conductance on one surface.
With \eqnref{eq:J_gen} and $\widetilde{A}_i=A_i+A^{pse}_i$, it is straightforward to derive {\eqJA} in the main text.

In the main text, the current density is shown only for the anti-parallel configuration.
In the parallel case, The two surface Hall conductances are the same ($\sigma_{b}=\sigma_t$)  and the total current density reads:
\eq{\label{eq:J_P}
J^\mu_{P}=\sigma_t\varepsilon^{\mu\rho\nu}\partial_\rho (\tilde{A}_{t,\nu}+\tilde{A}_{b,\nu})\ .
}
With the specific electromagnetic field configuration chosen in the main text, the above equation along $x$ can be further simplified into
\eq{
\label{eq:J_P_M}
J^x_P=-\frac{\sigma_t g_M (\dot{M}_{t,x}-\dot{M}_{b,x})}{v_f e}\ .
}
Clearly, $J^x_P=0$ when $M_{t,x}=M_{b,x}$.

On the other hand, the current induced by magnetic resonance along the $y$ direction in the anti-parallel configuration has the form
\eq{
\label{eq:J_yM}
J^y_{AP,M}=-\si_t \frac{1}{e v_f}g_M (\dot{M}_{t,y}+\dot{M}_{b,y})\ ,
}
which will also be used later.

\section{Details on Ferromagnetic Resonance}
In this part, we explicitly solve the Landau-Lifshitz-Gilbert (LLG) equation for the ferromagnetic resonance (FMR), which reads\cite{Kittel1948FMR,Kittel1976SSP,Aharoni2000FM}:
\begin{equation}
\label{eq:FM}
\dot{\bsl{M}}=-\gamma_0 \bsl{M}\times ( \bsl{B}_{eff}-\eta \frac{d \bsl{M}}{d t})\ ,
\end{equation}
where
$\bsl{M}$ is the magnetic moment,
$\bsl{B}_{eff}=\bsl{B}-\bsl{M}_N-\frac{K M_z}{M_s^2} \bsl{e}_z$,
$\bsl{B}$ is the applied magnetic field,
$M_s=|\bsl{M}|$,
$\gamma_0=g e/(2 m_e c)$ is the magneto-mechanical ratio with $g$ the Land\'e factor,
$\eta$ term is the Gilbert damping term,
$K$ is the anisotropy effect,
and $\bsl{M}_N=(N_x M_x, N_y M_y, N_z M_z)$ is the demagnetizing field with $N_x=N_y=0$ as the FMI is a film perpendicular to $z$ direction.
The equation can be rewritten as
\begin{equation}
\label{eq:FM_alt}
\dot{\bsl{M}}=-\gamma \bsl{M}\times \bsl{B}_{eff}-\frac{\gamma \alpha}{M_s} \bsl{M}\times(\bsl{M}\times \bsl{B}_{eff})\ ,
\end{equation}
where
$\alpha=\gamma_0 \eta M_s$ is the dimensionless damping constant with $|\alpha|\ll 1$, the sign of $\alpha$ is always chosen to make sure that the equation is damped, and $\gamma=\gamma_0/(1+\alpha^2)$.
\eqnref{eq:FM_alt} indicates that $\dot{M}_s=0$, and the constant magnitude indicates that only the motion of the direction $\bsl{n}=\bsl{M}/M_s$ is time-dependent and governed by \eqnref{eq:FM_alt}.
Since the in-plane magnetic field and magnetization are much smaller than $M_s$, i.e. $|n_x|\sim|n_y|\sim|B_{x}/M_s|\sim|B_{y}/M_s|\ll 1$, we can only keep terms up to the first order of those small quantities. \cite{Kittel1948FMR}
As a result, \eqnref{eq:FM_alt} is simplified to be
\eq{
\label{eq:FM_app}
\mat{
\frac{d}{d t}+\alpha \bar{\omega} & \bar{\omega} n_z\\
-\bar{\omega} n_z & \frac{d}{d t}+\alpha\bar{\omega}
}
\mat{
n_x \\
n_y
}
=
G_1
\mat{
B_x \\
B_y
}\ ,
}
where
$n_z=\sgn{M_z}$ is constant,
\eq{
G_1=
\mat{
\gamma \alpha & \gamma n_z \\
-\gamma n_z & \gamma \alpha
}\ ,
}
and $\bar{\omega}=\gamma B_{eff,z}n_z$.
Since the damping is present, we should use the Laplace transformation $\widetilde{n}_i(s)=\mathcal{L}[n_i](s)=\int^{+\infty}_0 dt e^{-s t} n_i(t)$, which gives $\mathcal{L}[\dot{n}_i](s)=s \widetilde{n}_i(s)-n_i(t=0)$.
With that, the \eqnref{eq:FM_app} is transformed to be
\eq{
\label{eq:FM_app_s}
G_0(s)
\mat{
\widetilde{n}_x(s)\\
\widetilde{n}_y(s)
}
-
\mat{
n_x(0)\\
n_y(0)
}
=
G_1
\mat{
\widetilde{B}_x(s)\\
\widetilde{B}_y(s)
}\ ,
}
where
\eq{
G_0(s)=
\mat{
s+\alpha \bar{\omega} & \bar{\omega} n_z\\
-\bar{\omega} n_z & s+\alpha\bar{\omega}
}
=U^{-1}
\mat{
s+(\alpha+\ii)\bar{\omega} & \\
 & s+(\alpha-\ii)\bar{\omega}
}
U\ ,
}
and $U=\exp[-\ii (\tau_x/2)(2\pi/4)n_z]$ with $\tau_x$ the Pauli matrix.
\eqnref{eq:FM_app_s} can directly give $\widetilde{n}_i(s)$, which in turn leads to $n_a(t)=\sum_j e^{s_j t} \text{Res}[\widetilde{n}_a(s_j)]$ with $j$ summing over all poles if $\widetilde{n}_i(s)$ only has isolated poles and decays fast enough at $|s|\rightarrow \infty$.

In the following, we give the steady solution of \eqnref{eq:FM_app} for $B_x=B_0 \cos(\omega t)$ and $B_y=0$.
In this case, $\widetilde{B}_x(s)=\frac{B_0}{2}(\frac{1}{s-\ii \omega}+\frac{1}{s+\ii \omega})$.
Combining the Laplace inverse transformation and \eqnref{eq:FM_app_s}, the steady solution has the form
\begin{equation}
\mat{
n_x \\
n_y
}
=
\frac{B_0}{2}
(e^{i\omega t} G_0^{-1}(i\omega)+e^{-i\omega t} G_0^{-1}(-i\omega))G_1
\mat{
1 \\
0
}
\ ,
\end{equation}
while the damped part decays as $e^{-\alpha \bar{\omega} t}$ with $\alpha \bar{\omega}>0$.
Since the steady solution has $e^{i \omega t}$ form, it can also be obtained with the Fourier transformation.
Explicitly, we have
\eqn{
&& n_x(t)=\frac{B_0 \gamma  \left(\alpha  \omega  \left(\left(\alpha ^2+1\right) \bar{\omega}^2+\omega ^2\right) \sin (t \omega )+\bar{\omega} \left(\left(\alpha ^2-1\right) \omega ^2+\left(\alpha ^2+1\right)^2 \bar{\omega}^2\right) \cos (t \omega )\right)}{2 \left(\alpha ^2-1\right) \omega ^2 \bar{\omega}^2+\left(\alpha ^2+1\right)^2 \bar{\omega}^4+\omega ^4} \nonumber\\
&& n_y(t)=\frac{B_0 \gamma  n_z \omega  \left(\left(\left(\alpha ^2+1\right) \bar{\omega}^2-\omega ^2\right) \sin (t \omega )-2 \alpha  \omega  \bar{\omega} \cos (t \omega )\right)}{2 \left(\alpha ^2-1\right) \omega ^2 \bar{\omega}^2+\left(\alpha ^2+1\right)^2 \bar{\omega}^4+\omega ^4} \ .
}
Since we only use $M_x=M_s n_x$ in the expression of the current, we only care the resonance of $n_x$, which is at
\eq{
\omega_r=\sqrt{\frac{\sqrt{\left(\alpha ^2+1\right)^2 \left(4 \alpha ^2+1\right)}-\left(\alpha ^2+1\right)^2}{\alpha ^2}} |\bar{\omega}|\ .
}
If only keeping the leading order of $\alpha$, then the expression of resonant $M_x$ become the same as that in the main text, and the resonant frequency is simplified to $\omega_r\approx \omega_0=|\gamma_0 (B_z n_z-N_z M_s-K/M_s)|$.


Since $n_y$ is proportional to $n_z$ when $B_z=0$, the FMR-induced current along the $y$ direction is zero in the anti-parallel case according to \eqnref{eq:J_yM}, if choosing the same approximation as the main text.

\section{Details on Anti-Ferromagnetic Resonance}

In this part, we solve the LLG equation for the anti-ferromagnetic resonance (AFMR), which reads
\begin{equation}
    \frac{d \bsl{M}_j}{dt}=-\gamma_0 \bsl{M}_j\times (\bsl{B}_{eff,j}-\eta \frac{d \bsl{M}_j}{dt})\ .
\end{equation}
Here $j=1,2$ indicates the two magnetic moments in one unit cell of anti-ferromagnetism, $M_s=|\bsl{M}_j|$ is constant and chosen to be independent of $j$, $\bsl{B}_{eff,j}=\bsl{B}+\bsl{B}_{E,j}+\bsl{B}_{A,j}$, $\bsl{B}$ is the applied magnetic field, $\bsl{B}_{E,1}=-\lambda \bsl{M}_2$ and $\bsl{B}_{E,2}=-\lambda \bsl{M}_1$ are the exchange fields, and $\bsl{B}_{A,j}=(-1)^{j-1} B_A \bsl{e}_z$ is the anisotropy effect.
Similar as the FMR case, we can rewrite the above equation into
\eq{
\label{eq:AFMR}
\frac{d\bsl{n}_j}{dt}=-\gamma \bsl{n}_j \times \bsl{B}_{eff,j}-\gamma \alpha \bsl{n}_j \times(\bsl{n}_j \times \bsl{B}_{eff,j})\ ,
}
where $\bsl{n}_j=\bsl{M}_j/M_s$, $\alpha=\gamma_0\eta M_s$ with $|\alpha|\ll 1$ is the damping constant, and $\gamma=\gamma_0/(1+\alpha^2)$.
Without loss of generality, we choose the magnetization to be in the $z$ direction with $\bsl{M}_1$ and $\bsl{M}_2$ pointing up and down, respectively, when the applied magnetic field  is zero.
In addition, we focus on the case where $B_x=B_0 \cos(\omega t)$ and $B_y=B_z=0$, and consider $|n_{j,x}|\sim |n_{j,y}|\sim |B_{x}/M_s|\sim |B_y/M_s| \ll 1$ so that the second and higher orders of them can be neglected.
In this case, \eqnref{eq:AFMR} is linearized as
\eq{
\frac{d}{dt}\mat{n_{1,x}\\ n_{2,x}\\ n_{1,y}\\ n_{2,y}}=G_0 \mat{n_{1,x}\\ n_{2,x}\\ n_{1,y}\\ n_{2,y}}+G_1 \mat{B_x\\ B_y}\ ,
}
where
\eq{
G_0=
\left(
\begin{array}{cccc}
 -B_{1} \alpha  \gamma  & -B_{E} \alpha  \gamma  & -B_{1} \gamma  & -B_{E} \gamma  \\
 -B_{E} \alpha  \gamma  & -B_{1} \alpha  \gamma  & B_{E} \gamma  & B_{1} \gamma  \\
 B_{1} \gamma  & B_{E} \gamma  & -B_{1} \alpha  \gamma  & -B_{E} \alpha  \gamma  \\
 -B_{E} \gamma  & -B_{1} \gamma  & -B_{E} \alpha  \gamma  & -B_{1} \alpha  \gamma  \\
\end{array}
\right)\ ,
}
\eq{
G_1=
\gamma  \left(
\begin{array}{cc}
 \alpha  & 1 \\
 \alpha  & -1 \\
 -1 & \alpha  \\
 1 & \alpha  \\
\end{array}
\right)\ ,
}
$B_1=B_E+B_A$ and $B_E=\lambda M_s$.
Using the same method as the FMR case, the steady solution of the above equation can be derived:
\eq{
\mat{n_{1,x}\\ n_{2,x}\\ n_{1,y}\\ n_{2,y}}
=
[e^{i\omega t} (i\omega-G_0)^{-1}+(-i\omega-G_0)^{-1} e^{-i\omega t}] G_1 \mat{B_0/2 \\ 0}\ .
}
This solution suggests that to the leading order of $\alpha$, the resonant frequency reads $\omega_1=\gamma_0\sqrt{B_1^2-B_E^2}=\gamma_0\sqrt{(2 B_E+B_A)B_A}$, and the steady solution along $x$ can be simplified as
\eq{
n_{1,x}=n_{2,x}=\frac{B_0\sqrt{B_1^2-B_E^2}}{2 \alpha B_1 (B_1 +B_E)}\sin(\omega_1 t)=\frac{B_0\gamma_0}{2 \alpha \omega_1}\frac{B_A}{B_A+B_E} \sin(\omega_1 t)\ .
}
With $M_{i,x}=M_s n_{i,x}$, {\eqAFMR} in the main text can be derived.

On the other hand, $n_{i,y}$ to the leading order of $\alpha$ reads
\eq{
n_{1,y}=-n_{2,y}=- \frac{B_0}{2 \alpha (B_A+B_E)}\cos(\omega_1 t)\ .
}
Clearly, $M_{t,y}$ and $M_{b,y}$ has opposite signs in the anti-parallel case, leading to zero AFMR-induced current to $O(1/\alpha)$ order according to \eqnref{eq:J_yM}.
\end{widetext}
\bibliography{bibfile_references}

\end{document}